\documentclass{aa}
\usepackage{epsfig,graphics}
\begin{document}

\title{The termination shock of a magnetar wind: a possible origin of gamma-ray burst X-ray afterglow emission}
\author{Z. Mao$^{1,2}$, Yun-Wei Yu$^{1,2,3}$, Z. G. Dai$^{4}$, C. M. Pi$^{5}$ and X. P.
Zheng$^{2}$}

\institute{$^1$Institute of Astrophysics, Huazhong Normal
University, Wuhan 430079, PR China \\
\email{yuyw@phy.ccnu.edu.cn}
\\$^2$Key Laboratory of Quark and Lepton Physics (Huazhong
Normal University), Ministry of Education, Wuhan 430079, PR China
\\$^3$Department of Physics, The University of Hong Kong, Hong Kong, PR China
\\$^4$Department of Astronomy, Nanjing University, Nanjing
210093, PR China
\\$^5$Department of Physics and Electronics, Hubei University of Education, Wuhan 430205, PR China}

\date{Received 7 September 2009 / Accepted 13 May 2010}
\abstract
{Swift observations suggest that the X-ray afterglow emission of
some gamma-ray bursts (GRB) may have internal origins, and the
conventional external shock (ES) cannot be the exclusive source of
the afterglow emission. }
{If the central compact objects of some GRBs are millisecond
magentars, the magnetar winds could play an important role in the
(internal) X-ray afterglow emission, which is our focus here.}
{The dynamics and the synchrotron radiation of the termination shock
(TS) of the magmnetar winds, as well as the simultaneous GRB ES, are
investigated by considering the magnetization of the winds.}
{As a result of the competition between the emission of the wind TS
and the GRB ES, two basic types of X-ray afterglows are predicted,
i.e., the TS-dominated and the ES-dominated types. Moreover, our
results also show that both of the two types of afterglows have a
shallow-decay phase and a normal-decay one, as observed by the
\textit{Swift} satellite. This indicates that some observed X-ray
afterglows could be (internally) produced by the magnetar winds, but
not necessarily GRB ESs.}
{} 

\titlerunning{GRB X-ray afterglows from magnetar winds}
\authorrunning{Z. Mao et al.}

\maketitle

\section{Introduction}
The conventional fireball model for gamma-ray bursts (GRBs; see
reviews by Piran 2005; M\'esz\'aros 2006) suggests that the burst
emission is produced by internal dissipations in a relativistic,
expanding fireball, while the afterglow emission arises from the
deceleration of the fireball in the circumburst medium by exciting
an external shock (ES). However, {\it Swift} observations have
showed that some X-ray afterglow emission components may sometimes
also have internal origins. The most solid evidence of this is the
rapid rise and decline of X-ray afterglow flares (e.g., Burrows et
al. 2005; Yu \& Dai 2009). Meanwhile, the ES model may also be
challenged by some plateau emission that is followed by an abrupt
cutoff (Troja et al. 2007). These phenomena indicate that internal
afterglow emission\footnote{In this paper we use the word ``internal
afterglow emission" to generally refer to the emission from any
sites inside the ES.} could be more usual than considered before, at
least in the X-ray band, whereas the ES is only one choice among
various afterglow origins. In this case, some observed chromatic
breaks in the afterglow light curves may also become more
understandable. The long-lasting internal emission strongly requires
long-lasting activities (energy release) of GRB central compact
objects after the bursts, and to a certain extent, spinning-down
magnetars are very plausible candidates (e.g., Dai et al. 2006;
Troja et al. 2007).

It has been proposed for nearly two decades that GRBs could be
produced by spinning-down, millisecond magnetars (Usov 1992;
Paczynski 1992; Duncan \& Thompson 1992). In this case, the
relativistic fireball responsible for a GRB could be associated with
a neutrino-driven wind (Thompson 1994; Metzger et al. 2007;
Bucciantini et al. 2007, 2009), with a magnetic
reconnection-accelerated wind (Drenkhahn 2002; Drenkhahn \& Spruit
2002; Giannois 2008), or with hyperaccretion onto the magnetars
(Zhang \& Dai 2008, 2009). Of more interest here, the postburst
magnetars could still keep abundant rotational energy, although a
considerable fraction of the energy has been expended on the bursts.
With the spin-down of the magnetars, this remaining energy can be
released and drive a continuous energy outflow, i.e., a postburst
wind. In view of the huge energy carried by the wind, Yu et al.
(2010) suggested that some observed X-ray afterglows could be
directly contributed by such winds, according to a qualitative
comparison between the model and the observations. However, in that
work, no specific radiation mechanism was addressed for the wind
emission besides introducing a free constant for the radiation
efficiency of the winds.

In this paper we try to search for a possible specific mechanism for
the wind emission. Following Dai (2004) and Yu \& Dai (2007), we
suggest that the termination shock (TS) of a magnetar wind, which
arises from the interaction of the wind with the GRB fireball and
the surrounding medium, may play an important role in the wind
emission. Based on a more careful consideration, we redescribe the
emission from the wind TS and the GRB ES by considering (1) the
existence of the TS after the spin-down timescale and (2) the
magnetization of the wind.

\section{The spin-down of a magnetar and the magnetar wind}
Some short-term and violent processes mean that a newly born
magnetar could drive an energetic, relativistic, complicated outflow
to produce a GRB. During the first tens of seconds after the birth
of the magnetar, its spin evolution must therefore be very
complicated. In contrast, on the relatively longer (afterglow)
timescales concerned here, the spin-down of the magnetar would be
mainly controlled by the electromagnetic torque (leading to an MHD
outflow) and the torque connected with gravitational wave radiation
(Zhang \& M\'esz\'aros 2001). For the latter, if the equatorial
ellipticity of the magnetar is mainly determined by the stellar
magnetic field, the energy losses through gravitational-quadrupole
radiation would not be essential for the spin-down of the star
unless the magnetic field is $>3\times10^{17}$ G (Usov 1992).

Therefore, for a magnetar with a magnetic field of $\sim10^{14-15}$
G, its postburst spin-down luminosity can be simply estimated by
(Shapiro \& Teukolsky 1983)
\begin{eqnarray}
L_{\rm
md}(t)&=&{(2\pi)^4\over6c^3}{ B_s^2R_s^6\over P_s^4}\nonumber\\
&\simeq& 10^{47}~{\rm erg~s^{-1}}R_{s,6}^6B_{s,14}^2P_{s,i,-3}^{-4}
\left(1+{t\over T_m}\right)^{-2},\label{lum}
\end{eqnarray}
where $R_{s,6}=R_{s}/10^6$ cm, $B_{s,14}=B_s/10^{14}$ G, and
$P_{s,i,-3}=P_{s,i}/1$ ms are the radius, magnetic field strength,
and initial spin period of the magnetar, respectively. The
characteristic spin-down timescale reads as
\begin{eqnarray}
T_{m}={6c^3\over(2\pi)^2}{I_sP_{s,i}^2\over
B_s^2R_s^6}\simeq2\times10^5~{\rm
s}~I_{s,45}R_{s,6}^{-6}B_{s,14}^{-2}P_{s,i,-3}^2,
\end{eqnarray}
where $I_{s,45}=I_s/10^{45}\rm~g~cm$ is the stellar moment of
inertial. The time $t$ is defined in the observer's frame, but the
time zero is actually not strictly set at the GRB trigger time since
Eq. (\ref{lum}) is only valid about $10-100$ s after the birth of
the magnetar. In other words, $t=0$ in this paper approximately
represents the transition time from the prompt phase to the
afterglow phase. Such a shift in the time zero would not influence
the calculation of the afterglow emission on much longer timescales
($t>10^{2-3}$ s).

With the release of the stellar rotational energy after the burst,
an MHD outflow is produced whose luminosity is determined by the
spin-down luminosity as shown in Eq. (\ref{lum}). Unlike the GRB
fireball that could be a neutrino-driven wind, this postburst
outflow is likely to be initially Poynting flux-dominated, mixed by
an electron-positron plasma. Since the fluctuating component of the
magnetic fields in the outflow can in principle be dissipated by
magnetic reconnection, the associated leptons are accelerated and
the outflow gradually becomes an ultrarelativistic kinetic-energy
flow consisting mainly of electron-positron plasma (lepton-dominated
wind; Coroniti 1990). For the bulk Lorentz factor of the accelerated
wind, we here arbitrarily take $\Gamma_{ w}\sim10^4$ as a fiducial
value following Atoyan (1999), who argues that the Crab pulsar wind
had Lorentz factor $\Gamma_{ w}\sim10^{4-7}$ to interpret the
measured radio spectrum of the Crab Nebula.

\section{The dynamics}
Owing to its ultrarelativistic velocity, the magnetar wind catches
up to and collide with the decelerating GRB ejecta. The interaction
of the wind with the ejecta and the surrounding medium gives rise to
a relativistic wind bubble, which includes four different regions
separated by a contact discontinuity surface and two shocks (i.e.,
the TS of the wind that propagates into the cold wind and the GRB ES
that propagates into the surrounding medium\footnote{The
initially-formed reverse and forward shocks in the GRB ejecta would
cross the ejecta very quickly ($\sim 100$ s), so we do not consider
the effect of these two shocks.}). To be specific, the four regions
are (1) unshocked medium, (2) shocked medium including the GRB
ejecta, (3) shocked wind, and (4) unshocked cold wind close to the
front of the TS. The shocked regions are emission regions.

We denote some quantities of region $i$ as follows: $n'_i$ is the
particle (proton or electron) number density, $B'_i$ the magnetic
field strength, $e'_i$ the internal energy density, $P'_i$ the
pressure, where the primes refer to comoving frame, and the
subscript $i$ represents regions $1-4$. $\Gamma_i$,
$\beta_i=(1-\Gamma_i^{-2})^{1/2}$, and $u_i=\beta_i\Gamma_i$ are the
bulk Lorentz factor, velocity, and four-velocity, respectively,
measured in the local medium's rest frame. It is convenient to
define a parameter
\begin{equation}
\sigma\equiv\frac{{B'}_4^2}{4\pi {n'}_4m_ec^2}
\end{equation}
to denote the degree of magnetization of the unshocked wind near the
front of the TS, where $m_e$ is the electron rest mass and $c$ the
speed of light. By assuming the wind is isotropic, the comoving
electron density of the unshocked wind can be expressed as
$n'_4=L_{\rm md}/[4\pi R^2\Gamma_4^2m_ec^3(1+\sigma)]$ with
$\Gamma_4=\Gamma_{w}$, and $R$ is the radius of the bubble in the
thin shell approximation\footnote{Following Eq. (7) in Dai (2004),
the thickness of the relativistic wind bubble,
$\Delta=R\left[\left(1+{1\over8\Gamma_2^2}\right)\left(1-{\chi\over1+8\Gamma_2^2}\right)-1\right]\ll
R$, could be negligible with respect to the bubble radius, where the
similarity variable $\chi$ is defined in Eq. (\ref{gamma3}).}. As
considered in Zhang \& Kobayashi (2004) and Fan et al. (2004), a
possible slow decrease in ${B'}_4^2$ at large radii due to magnetic
reconnection can be ignored. Thus same as $n'_4$, ${B'}_4^2$ would
approximatively decrease as $R^{-2}$ with the expansion of the
bubble, and thus the parameter $\sigma$ can be regarded as a
constant.

To connect the properties of the two sides of the shocks, some shock
jump conditions need to be listed as (Kennel \& Coroniti 1984; Zhang
\& Kobayashi 2004)
\begin{eqnarray}
\frac{e'_2}{n'_2m_pc^2}&=&\Gamma_2-1,\\
\frac{n'_2}{n'_1}&=&4\Gamma_2+3,\\
\frac{e'_3}{n'_3m_ec^2}&=&(\Gamma_{34}-1)f_a,\\
\frac{n'_3}{n'_4}&=&(4\Gamma_{34}+3)f_b,\\
{B'_3\over B'_4}&=&(4\Gamma_{34}+3)f_b, ~~\rm for ~\sigma>0
\end{eqnarray}
where an adiabatic index of $4/3$ is adopted, $m_p$ is the proton
rest mass,
$\Gamma_{34}=\left({\Gamma_3/\Gamma_4}+{\Gamma_4/\Gamma_3}\right)/2\approx{\Gamma_4/(2\Gamma_3)}$
is the bulk Lorentz factor of region 3 measured in the rest frame of
region 4,
\begin{eqnarray}
 f_a&=&1-\frac{(\Gamma_{34}+1)\sigma}
{2[\Gamma_{34}u_{3t}^2+(\Gamma_{34}^2-1)^{1/2}u_{3t}(u_{3t}^2+1)^{1/2}]},\nonumber\\
\\
f_b&=&{\Gamma_{34}u_{3t}+(\Gamma_{34}^2-1)^{1/2}(u_{3t}^2+1)^{1/2}\over{(4\Gamma_{34}+3)u_{3t}}}\label{fb1},
\end{eqnarray}
and $u_{3t}$ is the four-velocity of region 3 measured in the rest
frame of the TS. When $\Gamma_{34}\gg1$, which is easily satisfied
in this model (Yu \& Dai 2007), Eqs. (9) and (10) can be simplified
as
\begin{eqnarray}
f_a&\approx&1-\frac{\sigma}{2\left[u_{3t}^2+u_{3t}(u_{3t}^2+1)^{1/2}\right]},\\
f_b&\approx&{1\over4}\left[{1+\left(1+{1\over
u_{3t}^2}\right)^{1/2}}\right],
\end{eqnarray}
and $u_{3t}$ can be solved analytically from
$8(\sigma+1)u_{3t}^4-(8\sigma^2+10\sigma+1)u_{3t}^2+\sigma^2=0$
(Zhang \& Kobayashi 2004).

Following Eq. (4), the total kinetic energy of region 2 can be
calculated as $E_{k,2}=(\Gamma_2-1)(m_{\rm ej}+m_{\rm
sw})c^2+\Gamma_2(\Gamma_2-1)m_{\rm sw}c^2$, where $m_{\rm ej}$ and
$m_{\rm sw}$ are the rest masses of the GRB ejecta and swept-up
medium, respectively. Energy conservation requires that any increase
in $E_{k,2}$ should be equal to the work done by region 3, i.e.,
\begin{equation}
dE_{k,2}=\delta W=4\pi R^2(P'_{\rm th,3}+P'_{B,3})dR,
\end{equation}
where the thermal and magnetic components of the pressure of region
3 can be respectively calculated by $P'_{\rm
th,3}={1\over3}e'_3=\frac{4}{3}\Gamma_{34}^2n'_4m_ec^2f_af_b $ and
$P'_{B,3}={{B'}_3^2}/({8\pi})=8\Gamma_{34}^2n'_4m_ec^2\sigma f_b^2$
according to the shock jump conditions. Substituting the expression
of $E_{k,2}$ into Eq. (13), we can obtain
\begin{equation}
\frac{d\Gamma_2}{dR}=\frac{4\pi R^2[(P'_{\rm
th,3}+P'_{B,3})/c^2-(\Gamma_2^2-1)n_1m_p]}{m_{\rm
ej}+2\Gamma_2m_{\rm sw}}.\label{gamma2}
\end{equation}
Meanwhile, the relationships between the properties of the two sides
of the contact discontinuity surface show (Blandford \& McKee 1976)
\begin{eqnarray}
\Gamma_3&=&\Gamma_2\chi^{-1/2},\label{gamma3}\\
P'_3&=&P'_{\rm th,3}+P'_{B,3}=P'_2\chi^{-17/12},
\end{eqnarray}
where the similarity variable can be solved to be
\begin{equation}
\chi=\left[\frac{L_{\rm md}(f_af_b+6\sigma f_b^2)}{16\pi
R^2\Gamma_2^4n_1m_pc^3(1+\sigma)}\right]^{-12/29}\label{chi}.
\end{equation}
For a closing dynamic equation set, we also introduce another two
equations in order to calculate the increasing masses of regions 2
and 3 as
\begin{eqnarray} \frac{dm_{\rm sw}}{dR}=4\pi
R^2n_1m_p\label{m2}
\end{eqnarray}
and (Dai \& Lu 2002)
\begin{eqnarray}
\frac{dm_3}{dR}=4\pi
R^2\left(\frac{\Gamma_{34}\beta_{34}}{\Gamma_3\beta_3}\right)n'_4m_e\label{m3}.
\end{eqnarray}
Finally, we would like to pointed out that, in Dai (2004) and Yu \&
Dai (2007), the authors only took the existence of the TS before
$T_m$ into account since the shallow decay afterglow phase during
$\sim10^{3-5}$ s was mainly concerned there. In contrast, here such
an artificial cutoff of the TS has been abandoned for a more general
investigation.

Combining the above dynamic equations with the relationships of
$dt=(1-\beta_3)dR/(\beta_3c)$ for region 3 and
$dt=(1-\beta_2)dR/(\beta_2c)$ for region 2, we can numerically
calculate the temporal evolution of the shock dynamics. As mentioned
in Sect. 2, since $t=0$ is set at the transition time from the
prompt phase to the afterglow phase, the initial conditions for our
calculation is taken as $R_{i}=10^{16}$ cm, $\Gamma_{2, i}=150$, and
$E_{k,2,i}=10^{51}$ erg, which correspond to the deceleration
timescale of the GRB ejecta. Figure 1 shows the evolution of the
bulk Lorentz factors of the wind TS and the GRB ES with different
values of $\sigma$. The nearly overlapping of the curves with
$\sigma$ varying from 0 to 3 indicates that the dynamic evolutions
of both the shocks are insensitive to the degree of magnetization of
the wind. It is not surprising to obtain such a result, which
actually has been know for a long time for ordinary pulsar wind
nebulae (e.g., Emmering \& Chevalier 1987; Bucciantini et al. 2003;
Del Zanna et al. 2004).

\section{The radiation from the shocks}
With the propagation of the shocks, the bulk kinetic energy of the
GRB ejecta and the magnetar wind would be gradually converted into
internal energy of the shocked materials. As usual, we assume that
the internal energy of the shocked medium is shared by magnetic
fields, electrons, and protons with fractions $\epsilon_{B}$,
$\epsilon_{e}\sim\sqrt{\epsilon_{B}}$ (Medvedev 2006) and
$1-\epsilon_{e}-\epsilon_{B}$, respectively. For the shocked wind,
the energy fractions of the leptons and magnetic fields can be
determined by the shock jump condition with a certain $\sigma$.
Considering the shock acceleration of charge particles, the
electrons in shocked regions are assumed to distribute as
$n_{{\gamma'},i}^{'} \propto {\gamma'}_i^{-p}$ with a minimum
electron Lorentz factor: ${\gamma'}_{m,2}
=\epsilon_{e}g_p(m_p/m_e)(\Gamma_2-1)$ for region 2 and
${\gamma'}_{m,3}=g_p{e'_3}/({n'_3m_ec^2})$ for region 3 with
$g_p\equiv(p-2)/(p-1)$. Owing to the synchrotron cooling of the
electrons, a cooling Lorentz factor ${\gamma'}_{c,i} =6\pi
m_ec/(\sigma_{_{T}}{B'}^2_i\Gamma_it)$ should also be defined by
equaling the cooling time to the dynamic time, where $\sigma_{_{T}}$
is Thomson cross section. Characterized by ${\gamma'}_{m}$,
${\gamma'}_{c}$ and $p$ (the subscript $i$ is omitted hereafter), a
quasi-static distribution of the electrons can be written as (Huang
et al. 2000)
\begin{equation}
{n_{{\gamma'}}^{'}\propto}\left\{
\begin{array}{ll}
\left(\frac{{\gamma'}}{{\gamma'}_{L}}\right)^{-x},
 &{\gamma'}_{L}\leq{\gamma'}\leq{\gamma'}_{H},\\
\left(\frac{{\gamma'}_{H}}{{\gamma'}_{L}}\right)^{-x}\left(\frac{{\gamma'}}{{\gamma'}_{H}}\right)^{-p-1},
 &{\gamma'}_{H}<{\gamma'}\leq \gamma'_{M},
\end{array}\right.
\end{equation}
where $\gamma'_{H}=\max(\gamma'_{m},\gamma'_{c})$,
$\gamma'_L=\min(\gamma'_m,\gamma'_{c})$, and $x=2$ for
${\gamma'}_{c}\leq{\gamma'}_{m}$ and $x=p$ for
${\gamma'}_{c}>{\gamma'}_{m}$. $\gamma'_M\sim q_eB'R/(m_ec^2)$ is
the maximum electron Lorentz factor with $q_e$ the electron charge.

For the X-ray band of interest in this paper, we only consider the
synchrotron radiation of the electrons in both regions 2 and 3. The
synchrotron emission coefficient at frequency $\nu'$ can be given by
(Rybicki \& Lightman 1979)
\begin{eqnarray}
{j'}_{\nu'}=\frac{1}{4\pi}\int_{\gamma'_L}^{\gamma'_M}
{n'}_{{\gamma'}}\left[\frac{\sqrt{3}q_e^3B'}{m_ec^2}\frac{{\nu'}}{{\nu'}_0}
\int_{\nu^{'}\over{\nu'}_0}^{\infty}K_{5\over3}(y)dy\right]d{\gamma'},
\end{eqnarray}
where ${\nu'}_0=3q_eB'{\gamma'}^2/(4\pi m_ec)$, $K_{5\over3}(y)$ is
the Bessel function. Then by assuming the viewing angle is zero, we
can calculate the synchrotron flux density received by the observers
as (Yu et al. 2007)
\begin{equation}
F_{\nu}(t)=\frac{1}{d_{L}^2}\int_0^{\pi/2}\frac{{j'}_{{\nu'}}V'}
{\Gamma^3(1-\beta\cos{\theta})^3}\frac{\sin{\theta}}{2}\cos{\theta}d\theta\label{fnu},
\end{equation}
where $d_{L}\sim10^{28}$ cm is the luminosity distance of the GRB
and $V'$ is the comoving volume of the emission region. Considering
the equal-arrival-time effect for a fixed observer's time $t$,
different angles $\theta$ should correspond to different radii $R$,
satisfying (Huang et al. 2000)
\begin{equation}
t=\int_0^R\frac{1-\beta(r)\cos{\theta}}{\beta(r)c}dr.
\end{equation}
Therefore, as functions of radius, $j'_{\nu'}$ and $V'$ should also
be functions of $\theta$ in the integral in Eq.(22).

Some X-ray ($\sim$1 keV) light curves contributed by the wind TS and
the GRB ES as well as the combination of them are shown in Fig. 2
with different values of $\sigma$ and $\epsilon_{e}$. These two
parameters control the energy densities of the electrons and
magnetic fields, and thus determine the radiation efficiency of the
two emission regions. Fig. 2 shows that the emission from the ES is
mainly dependent on the value of $\epsilon_{e}$ (upper panel) but
insensitive to $\sigma$ (lower panel). This is because the variation
in $\sigma$ has little influence on the shock dynamics. In contrast,
the lower panel of Fig. 2 shows that the emission from the TS is
strongly dependent on the value of $\sigma$. More specifically,
because a weak magnetic field can suppress the synchrotron radiation
and a strong magnetic field reduces the energy fraction of the
electrons, neither too low nor too high $\sigma$ leads to strong TS
emission, which can be seen clearly in Fig. 3. For the model
parameters adopted in Fig.3, $\sigma=\sigma^*\approx0.05$ results in
the strongest TS emission. When the value of $\sigma$ is much closer
to $\sigma^*$ or $\epsilon_e$ is sufficiently small (e.g.,
$\epsilon_e\sim0.02$), the TS emission is very easy to exceed the ES
emission. As a result, three types of X-ray afterglows can be
predicted, i.e., the TS-dominated type, the ES-dominated type and
the intermediate type, as shown by the left, right, and middle
panels of Fig. 2, respectively. Qualitatively, such a result is
basically consistent with the conjecture proposed by Yu et al.
(2010).

As found previously (e.g., Dai 1998a,b; Zhang \& M\'esz\'aros 2001),
the profile of the ES-dominated afterglow light curves (e.g., the
cases in the right panel of Fig. 2) is in good agreement with the
observational light curves including a shallow decay phase and a
normal decay phase (Zhang et al. 2006), because of the energy
injection into the ES before $T_m$. More surprisingly, on the other
hand, we find that a very similar profile also appears in the
TS-dominated cases (e.g., the cases in the left panel of Fig. 2).
Please note that the ``normal decay" here is actually not produced
by the ES. To understand such an unexpected ``normal decay", we
adopt an approximative dynamic evolution of the TS as
$\Gamma_3\propto t^{-1/2}$ for $t>T_m$, which is extracted from the
numerical results shown in Fig. 1. Combining with $n'_4\propto
L_{\rm md}/(\Gamma_3^2t)\propto t^{-2}$, we can get $B'_3\propto
t^{-1/2}$, $\gamma'_{m,3}\propto t^{1/2}$, and $\gamma'_{c,3}\propto
t^{1/2}$. Then following Sari et al. (1998), we can easily derive
$F_{\nu}\propto t^{-1}$ from the characteristic spectral quantities
$\nu_m\propto t^0$, $\nu_c\propto t^0$, $F_{\nu,\max}\propto
t^{-1}$, and the relationship $\nu_m<\nu_X<\nu_c$. The ratio of
$F_{\nu}/L_{\rm md}\propto t$ indicates that the X-ray luminosity of
the wind emission does not track the spin-down luminosity of the
magnetar with a constant fraction, which is assumed in Yu et al.
(2010). However, the more important conjecture proposed by Yu et al.
(2010) that some observed X-ray afterglows are directly produced by
the magnetar winds can be still strongly favored by the good
agreement between the profiles of the TS-dominated afterglow light
curves and the observational ones.

Finally, although both the ES- and TS-dominated afterglow light
curves are qualitatively consistent with the observations, some tiny
differences between them could also be found by some careful
statistical studies. For example, the shallow decay phase of the
TS-dominated afterglows would be slightly flatter than for the
ES-dominated afterglows, and the same is true for the so-called
normal decay phase.

\section{Summary}
Following Dai (2004) and Yu \& Dai (2007), we investigated the
emission from a TS of a magnetar wind both before and after $T_m$ as
well as the simultaneous emission from the GRB ES, where the
magnetization of the wind was taken into account. The competition
between the emission of the two shocks is mainly determined by the
model parameters $\sigma$ and $\epsilon_{e}$. As a result, an
ES-dominated type and a TS-dominated type of afterglows are
predicted, both of which are roughly consistent with the
observational X-ray afterglows. This indicates that some observed
afterglows could come from the wind TS, but not necessarily from the
ES. To a certain extent, we regard such TS emission as general
internal afterglow emission, although the radius of the TS is nearly
as large as that of the ES. To distinguish between these two types
of afterglows, some tiny differences in the temporal indices could
be useful for some statistical studies such as that in Yu et al.
(2010). In this paper, very high values of $\sigma$ were not
considered, because in those cases the TS may disappear (Giannios et
al. 2008) and the injecting flow is probably dominated by Poynting
flux. This was studied thoroughly in Dai \& Lu(1998a,b) and Zhang \&
M\'esz\'aros (2001).

\begin{acknowledgements}
This work is supported by the opening project of the Key Laboratory
of Quark and Lepton Physics (MOE) at CCNU (grant no. QLPL2009P01)
and by the Self-Determined Research Funds of CCNU (grant no.
CCNU09A01020) from the colleges' basic research and operation of MOE
of China. X. P. Zheng is supported by the National Natural Science
Foundation of China (grant no. 10773004).
\end{acknowledgements}

\begin{figure}
\resizebox{\hsize}{!}{\includegraphics{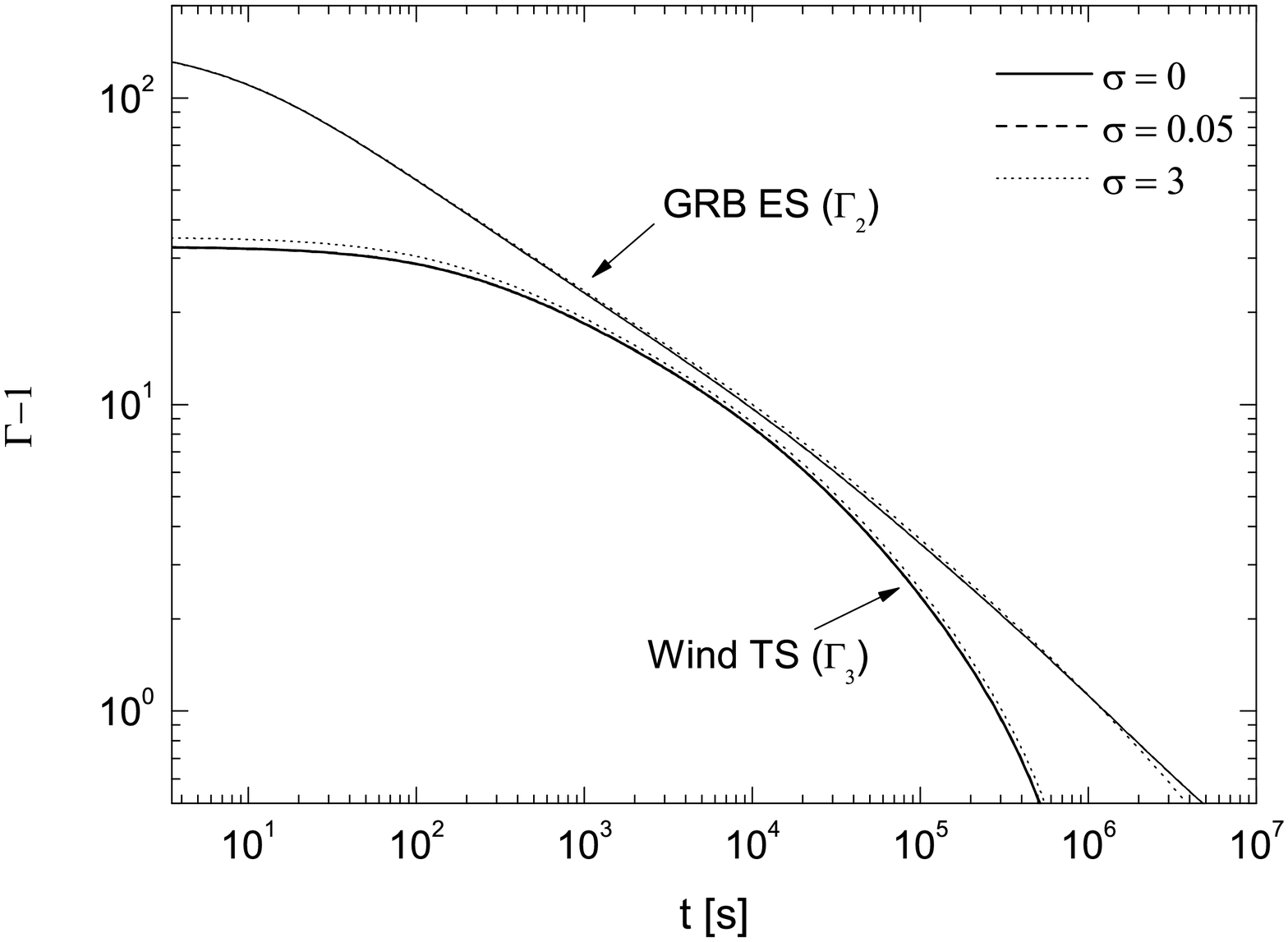}} \caption{The
evolution of the bulk Lorentz factors of the GRB ES ($\Gamma_{2}$)
and the wind TS ($\Gamma_{3}$) with different values of $\sigma$ as
labeled. The parameters for the magnetar are as follows:
$R_{s,6}=1$, $P_{s,i,-3}=3$, $I_{s,45}=1$, $B_{s,14}=10$.}
\end{figure}

\begin{figure}
\resizebox{\hsize}{!}{\includegraphics{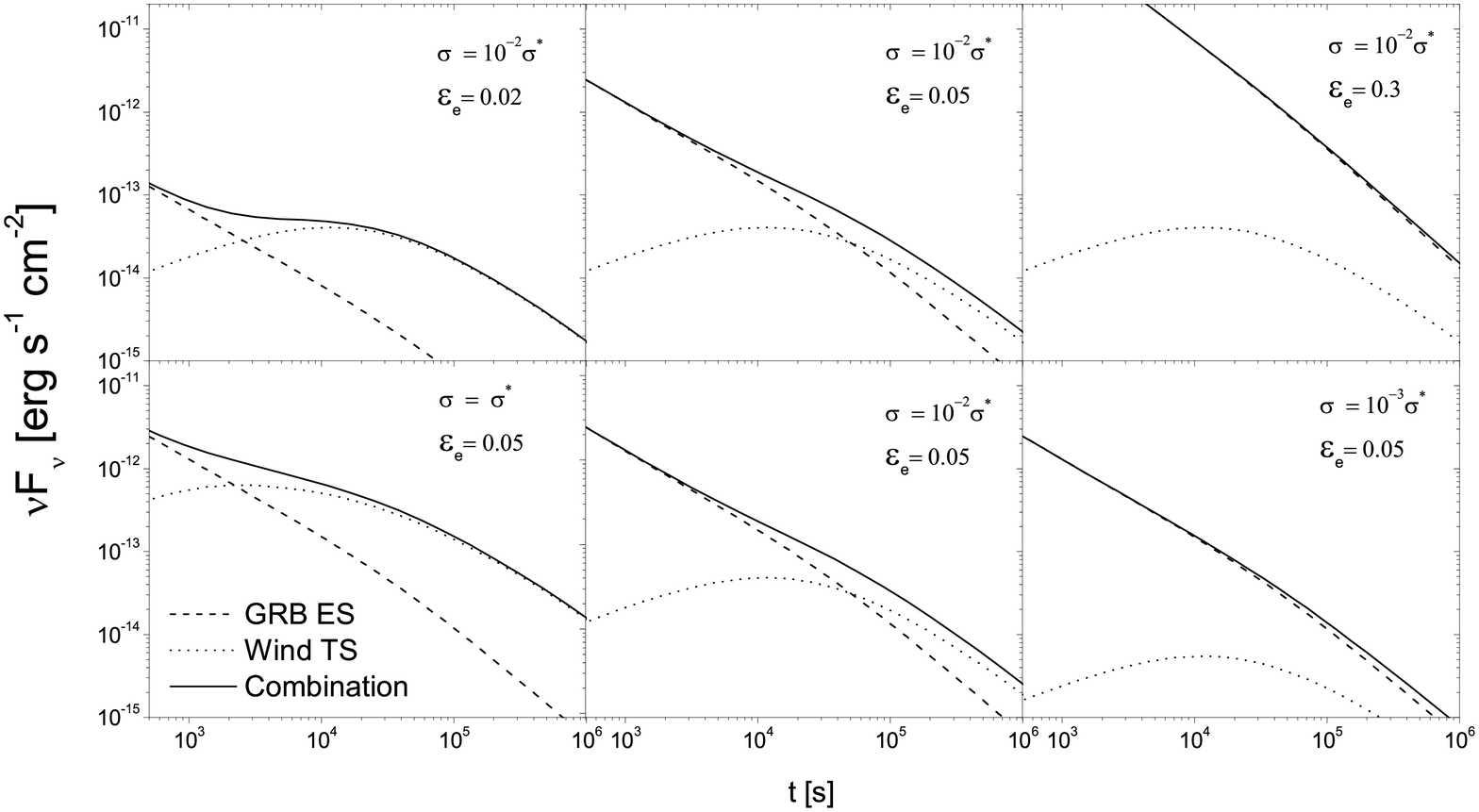}} \caption{X-ray
(1keV) light curves of the GRB ES (dashed lines) and the wind TS
(dotted lines) as well as their combination (solid lines). The
values of $\epsilon_{e}$ and $\sigma$ are as labeled and $p=2.5$.
The other model parameters are the same as those in Fig. 1.}
\end{figure}
\begin{figure}
\resizebox{\hsize}{!}{\includegraphics{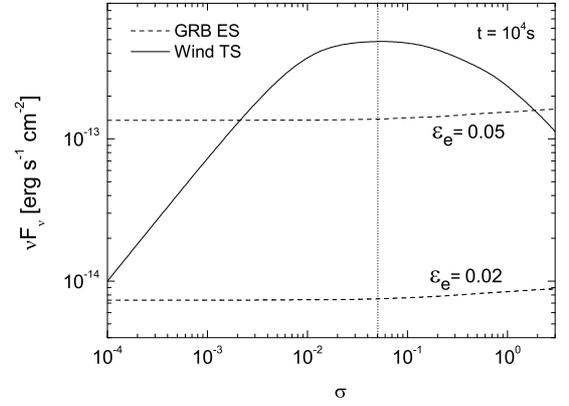}} \caption{The
$\sigma$-dependence of the synchrotron flux of the GRB ES (dashed
line) and the wind TS (solid line) at the time of $10^4$ s. The
vertical dotted line represents $\sigma^*=0.05$. The model
parameters are the same as those in Fig. 2.}
\end{figure}

\end{document}